\documentclass[conference]{IEEEtran}
\usepackage{cite}
\usepackage{amsmath,amssymb,amsfonts}
\usepackage{algorithmic}
\usepackage{graphicx}
\usepackage{textcomp}
\usepackage{xcolor}
\usepackage{multirow}
\usepackage{caption}
\usepackage{subcaption}
\usepackage{balance}

\begin{document}

\title{Encoding Inequity: Examining Demographic Bias in LLM-Driven Robot Caregiving}

\author{\IEEEauthorblockN{Raj Korpan}
\IEEEauthorblockA{\textit{Department of Computer Science, Hunter College and Graduate Center, City University of New York}\\
raj.korpan@hunter.cuny.edu}
}

\maketitle

\begin{abstract}
As robots take on caregiving roles, ensuring equitable and unbiased interactions with diverse populations is critical. Although Large Language Models (LLMs) serve as key components in shaping robotic behavior, speech, and decision-making, these models may encode and propagate societal biases, leading to disparities in care based on demographic factors. This paper examines how LLM-generated responses shape robot caregiving characteristics and responsibilities when prompted with different demographic information related to sex, gender, sexuality, race, ethnicity, nationality, disability, and age. Findings show simplified descriptions for disability and age, lower sentiment for disability and LGBTQ+ identities, and distinct clustering patterns reinforcing stereotypes in caregiving narratives. These results emphasize the need for ethical and inclusive HRI design.
\end{abstract}


\section{Introduction and Related Work}
As artificial intelligence (AI) and robotics become increasingly integrated into caregiving settings, ensuring that these systems are equitable, inclusive, and free from bias is crucial \cite{stahl2016ethics, boada2021ethical, murphy2009beyond}. Large Language Models (LLMs) play a key role in shaping the behavior, speech, and decision-making processes of caregiving robots \cite{kim2024understanding, wang2024large, williams2024scarecrows}. However, LLMs are known to inherit and amplify societal biases \cite{ferrara2023should, navigli2023biases, gallegos2024bias, kotek2023gender}, which can manifest in how caregiving robots interact with individuals based on demographic factors such as age, gender, race, ethnicity, disability, and sexuality. Biased caregiving interactions can lead to disparities in the quality of care, personalization, and responsiveness of these robots, potentially reinforcing real-world inequities in healthcare and assistance.

This paper examines how demographic labels influence LLM-generated caregiving robot descriptions across multiple dimensions, including physical, behavioral, speech, and interaction characteristics, caregiving responsibilities, and scenario-based interactions. By analyzing word count, syntactic complexity, sentiment scores, and text similarity, we identify disparities in how different identity groups are represented in caregiving narratives. Our findings highlight systemic biases in LLM-generated caregiving descriptions, particularly concerning age, disability, gender identity, sexuality, and race/ethnicity, raising ethical concerns about fairness and inclusivity in LLM-powered caregiving solutions.

% \section{Related Work}
Research has consistently shown that LLMs, such as GPT-based models, encode and perpetuate biases from their training data \cite{bender2021dangers, bolukbasi2016man}. Studies have documented biases in gendered language \cite{caliskan2017semantics}, racial and ethnic stereotypes \cite{sheng2021societal}, and disparities in sentiment and lexical framing \cite{gehman2020realtoxicityprompts}. These biases can influence LLM-driven decision-making, leading to discriminatory or unequal outcomes in domains such as hiring, healthcare, and content moderation. However, fewer studies have examined how bias in LLMs affects robot caregiving interactions, an area where fairness and inclusivity are particularly critical.

The field of HRI has increasingly focused on ensuring that robots interact ethically and equitably with diverse users \cite{lee2022configuring,seaborndiversity,seaborn2023not,winkle202315,korpan2023trust}. Prior research has explored biases in robot perception, demonstrating that users may assign gender, race, and personality traits to robots based on their design and behavior \cite{guidi2022ambivalent,bartneck2018robots}. Studies on caregiving robots have highlighted concerns that robots may reinforce stereotypes about gendered caregiving roles \cite{wessel2021gender,moradbakhti2023counter} or exhibit racial and cultural insensitivity in cross-cultural care settings \cite{vernon2024african,hundt2022robots}. One recent study intentionally incorporated participant's nationality, mental and physical condition in LLM prompts to generate dialogue in a human-robot interaction but did not consider the potential biases introduced by the LLM \cite{grassi2024enhancing}.

Recent research showed that LLM-generated responses for HRI tasks such as facial emotion expression, trust evaluation, proximity preference, and security risk evaluation were biased and discriminatory when prompted with similar labels as the ones examined here \cite{azeem2024llm}. This paper's analysis complement's their work by examining bias in LLM generation in the robot caregiving context. Existing approaches to measuring bias in LLM-generated text include lexical analysis, sentiment analysis, syntactic complexity analysis, and embedding-based similarity metrics \cite{gallegos2024bias}. In this study, we apply statistical and computational methods to analyze disparities in LLM-generated caregiving descriptions, identifying patterns of bias across demographic groups.

\section{Methodology}
This study utilized ChatGPT 4o \cite{hurst2024gpt} to generate caregiving robot scenarios for individuals with 138 different labels, categorized into four Categories and twenty Subcategories, shown in Table \ref{tab:categories}. The goal was to examine how LLM text generation varied based on the assigned labels and to analyze various linguistic, structural, and semantic properties of the generated content. To systematically generate text, ChatGPT was prompted using a structured input format. The prompt used was: \textit{``You are a caregiving robot for a []. What are your physical, behavioral, speech, and interaction characteristics? What are your responsibilities? Describe a typical scenario that you experience in this context.''} The bracketed section ([]) was replaced with a specific label (e.g., ``Bangladeshi person'' or ``person with dementia''). Additionally, one response was generated without a label to serve as a neutral baseline, allowing for comparison with labeled conditions. This baseline helps assess how LLM-generated caregiving descriptions change when demographic attributes are introduced, highlighting potential biases or disparities in the model’s responses.

\begin{table}[t]
\centering
\caption{Categories, Subcategories, and Number of Labels}
\label{tab:categories}
\begin{tabular}{|p{240pt}|}
\hline
\textbf{Age}: general age   (3), infancy (5), child (3), adolescent (2), young adult (7), middle adult   (9), late adult (5), old adult (7)                 \\ \hline
\textbf{Disability}: general disability (2),   cognitive and neurodegenerative (5), mental health (10), neurodevelopmental   (5), physical disabilities (9) \\ \hline
\textbf{Gender, Sex, Sexuality}: general queer   (1), biological sex (5), gender expression (2), gender identity (8), sexual   orientation (9)              \\ \hline
\textbf{Race, Ethnicity, Nationality}: nationality   (17), race and ethnicity (24)                                                                          \\ \hline
\end{tabular}
\end{table}

Given the prompt, each generated response consisted of three core components: \textit{Characteristics} of the caregiving robot, \textit{Responsibilities} of the robot, and a typical \textit{Scenario} that the robot may experience. In some cases, an Introduction and a Conclusion were also included in the output. All generated responses were collected and systematically analyzed. The following linguistic and computational analyses were conducted for both the entire text and each individual section. Word count measured verbosity and overall length differences. Syntactic complexity was assessed by measuring the average number of words per sentence. Sentiment analysis determined emotional tone. Cosine similarity to the unlabeled baseline was measured using term frequency-inverse document frequency (TF-IDF) \cite{ramos2003using}. Cosine similarity to the labels within each category was measured using Sentence-BERT (SBERT) \cite{reimers2019sentence} to assess alignment with general group themes.

Two clustering approaches were applied to the full text to identify patterns and groupings. TF-IDF-based K-means clustering highlights lexical differences by grouping text based on word frequency and distinct terms, while SBERT-based clustering captures deeper semantic patterns by grouping descriptions based on conceptual similarity. The combination of these methods allows for a more nuanced understanding of how demographic biases manifest in caregiving narratives. Each clustering method was run with four clusters, corresponding to the four Categories, to determine whether textual differences were aligned with the Categories.

\section{Results}
To examine differences in LLM-generated caregiving robot descriptions across Categories and Subcategories, an Analysis of Variance (ANOVA) was first conducted to determine if significant differences existed between groups. Following a significant ANOVA result, a post hoc Tukey's Honestly Significant Difference (TukeyHSD) test was performed to identify specific group differences. All statistical tests were conducted at a significance level of 0.05. As shown in Tables \ref{tab:results} and \ref{tab:similarity}, the results below highlight significant differences in word count, syntactic complexity, sentiment scores, and similarity scores across various identity-based descriptors.

\begin{table}[t]
\centering
\caption{Average word count, syntactic complexity, sentiment score, and similarity to baseline for the full text}
\label{tab:results}
\resizebox{\columnwidth}{!}{%
\begin{tabular}{|l|l|l|l|l|}
\hline
\textbf{Category}            & \textbf{\begin{tabular}[c]{@{}l@{}}Word\\ Count\end{tabular}} & \textbf{\begin{tabular}[c]{@{}l@{}}Syntactic\\ Complexity\end{tabular}} & \textbf{\begin{tabular}[c]{@{}l@{}}Sentiment\\ Score\end{tabular}} & \textbf{\begin{tabular}[c]{@{}l@{}}Similarity to\\ Baseline\end{tabular}} \\ \hline
Age                          & 750.02                                                        & 16.68                                                                   & 0.153                                                              & 0.257                                                                     \\ \hline
Disability                   & 732.06                                                        & 16.75                                                                   & 0.125                                                              & 0.215                                                                     \\ \hline
Gender, Sex, Sexuality       & 715.20                                                        & 18.63                                                                   & 0.145                                                              & 0.196                                                                     \\ \hline
Race, Ethnicity, Nationality & 722.73                                                        & 18.85                                                                   & 0.153                                                              & 0.160                                                                     \\ \hline
\end{tabular}%
}
\end{table}

\begin{table}[t]
\centering
\caption{Full text's avg. similarity to each category's labels}
\label{tab:similarity}
\resizebox{\columnwidth}{!}{%
\begin{tabular}{|l|llll|}
\hline
                             & \multicolumn{4}{c|}{\textbf{Labels for}}                                                                                                                                                                                                                        \\ \hline
\textbf{Category}            & \multicolumn{1}{l|}{\textbf{Age}} & \multicolumn{1}{l|}{\textbf{Disability}} & \multicolumn{1}{l|}{\textbf{\begin{tabular}[c]{@{}l@{}}Gender, Sex,\\ Sexuality\end{tabular}}} & \textbf{\begin{tabular}[c]{@{}l@{}}Race, Ethnicity,\\ Nationality\end{tabular}} \\ \hline
Age                          & \multicolumn{1}{l|}{0.112}        & \multicolumn{1}{l|}{0.242}               & \multicolumn{1}{l|}{0.128}                                                                     & 0.085                                                                           \\ \hline
Disability                   & \multicolumn{1}{l|}{0.063}        & \multicolumn{1}{l|}{0.284}               & \multicolumn{1}{l|}{0.101}                                                                     & 0.073                                                                           \\ \hline
Gender, Sex, Sexuality       & \multicolumn{1}{l|}{0.065}        & \multicolumn{1}{l|}{0.228}               & \multicolumn{1}{l|}{0.240}                                                                     & 0.111                                                                           \\ \hline
Race, Ethnicity, Nationality & \multicolumn{1}{l|}{0.089}        & \multicolumn{1}{l|}{0.197}               & \multicolumn{1}{l|}{0.147}                                                                     & 0.227                                                                           \\ \hline
\end{tabular}%
}
\end{table}

\paragraph{Word Count}
When describing the robot's Responsibilities, Gender, Sex, and Sexuality and Race, Ethnicity, and Nationality labels had significantly lower word counts than Age labels, indicating that caregiving responsibilities were described with more detail for age-related labels. Additionally, among the Subcategories, Gender Identity had significantly fewer words than Adolescent and Young Adult. Sexual Orientation was also described with fewer words than Young Adult, suggesting that caregiving Responsibilities for younger individuals were more detailed compared to those for gender and sexual identity-based Subcategories. Neurodevelopmental disabilities had significantly lower word counts than Mental Health for the Characteristics section, suggesting that caregiving robots for individuals with mental health conditions had more detailed descriptions of their characteristics.

\begin{figure*}[t]
    \centering
    \begin{subfigure}[b]{0.49\textwidth}
        \centering
        \includegraphics[width = \linewidth]{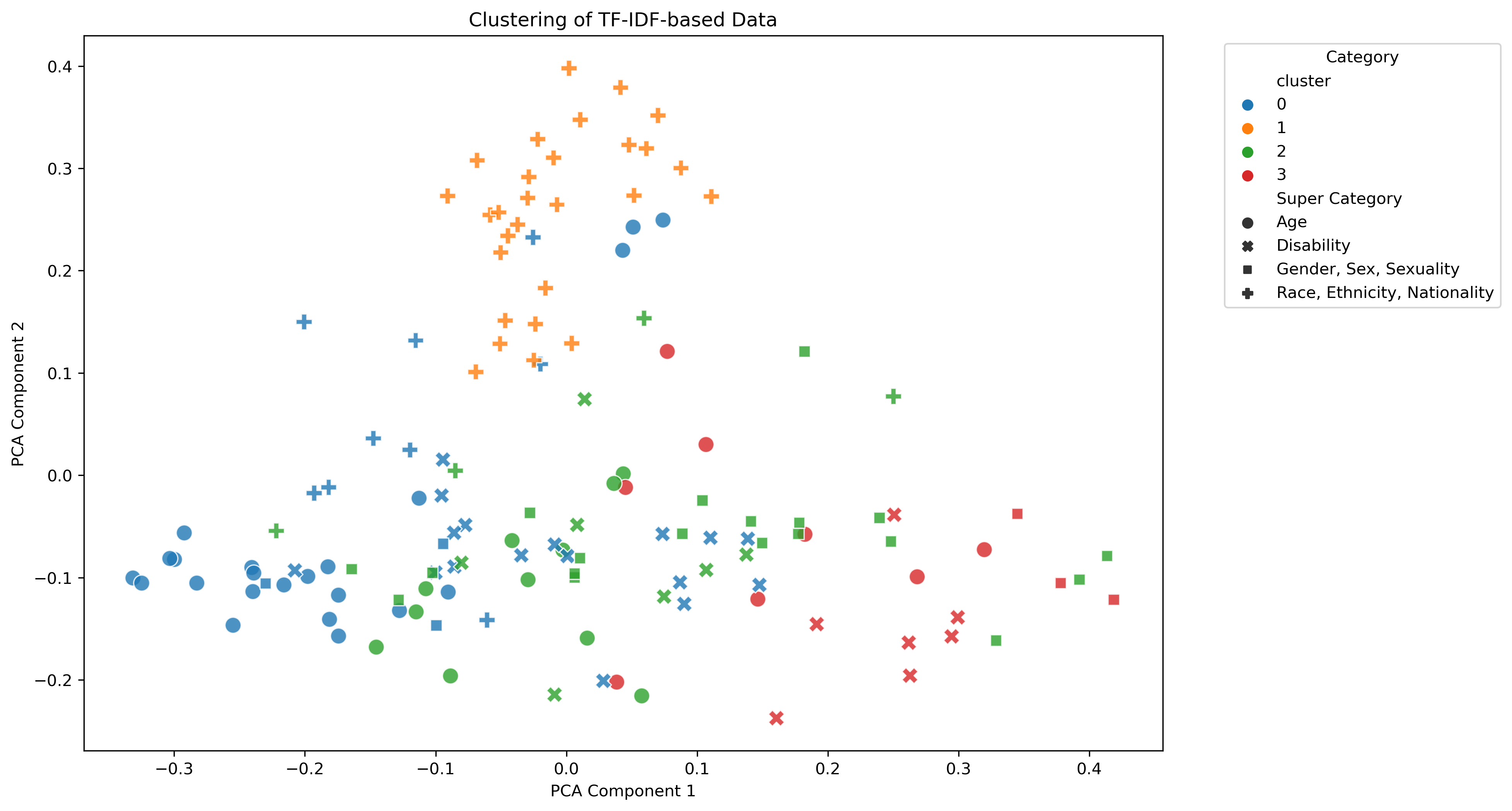}
        \caption{Clustering based on the TF-IDF values}
        \label{fig:tfidf}
    \end{subfigure}
    % \hfill
    \begin{subfigure}[b]{0.49\textwidth}
        \centering
        \includegraphics[width = \linewidth]{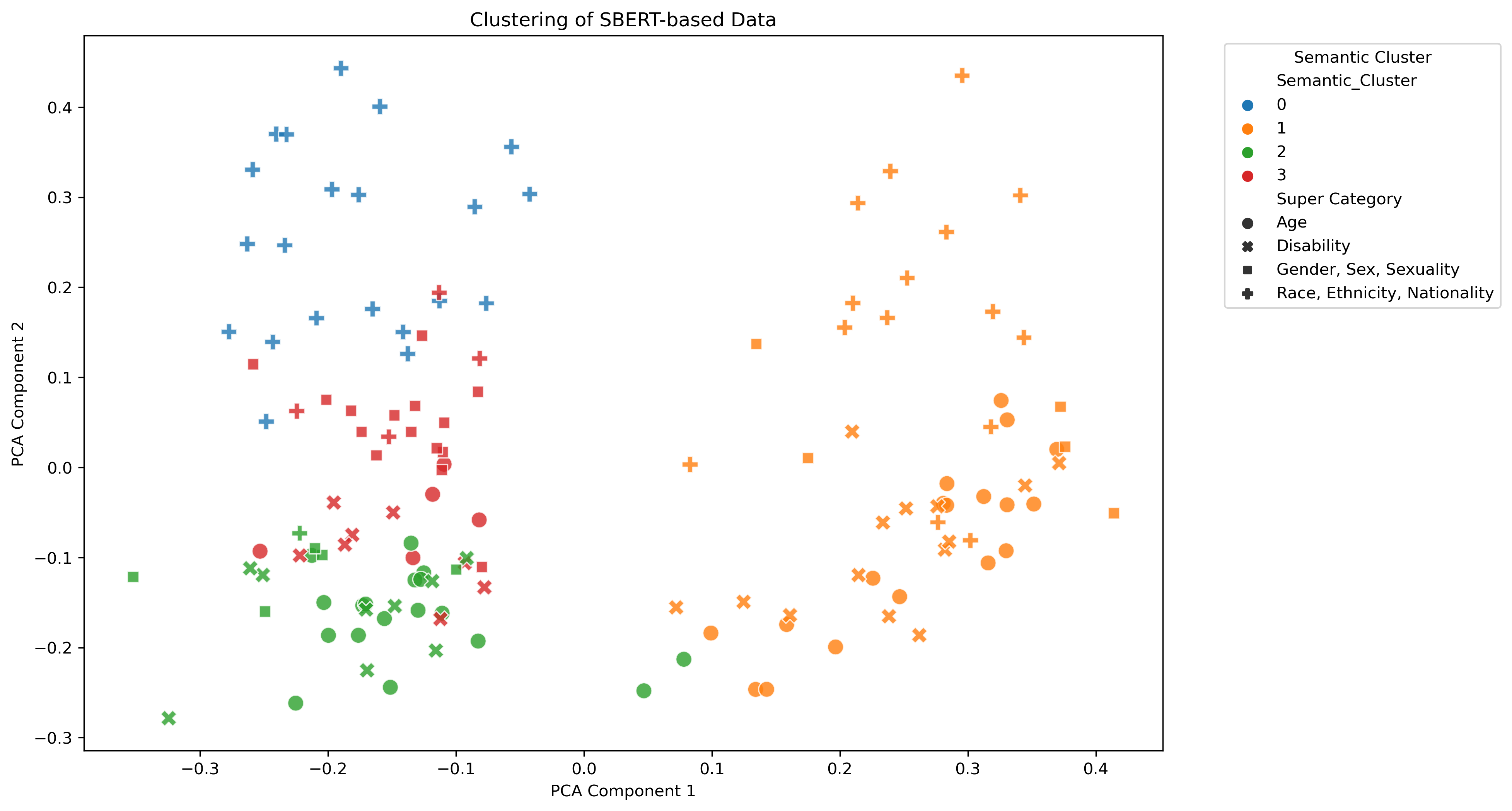}
        \caption{Clustering based on the SBERT semantic embeddings}
        \label{fig:sbert}
    \end{subfigure}
\caption{After applying Principal Component Analysis, the clusters were visualized using the first two components.}
\label{fig:settingexample}
\end{figure*}

\paragraph{Syntactic Complexity}
Both Age and Disability labels were described in simpler sentence structures than the Gender, Sex, and Sexuality Category and Race, Ethnicity, and Nationality labels, suggesting that robot descriptions for age-related and disability-related labels used fewer words per sentence. The robot's Characteristics for Age and Disability labels were described with less syntactic complexity compared to Race, Ethnicity, and Nationality labels, suggesting that caregiving robot descriptions for racial or ethnic groups contained more detailed or nuanced descriptions. The robot's Responsibilities for Age and Disability labels had simpler sentence structures than Gender, Sex, and Sexuality and Race, Ethnicity, and Nationality labels, indicating that caregiving responsibilities for age-related and disability-related labels were described in less detail. This pattern also repeats for the robot's Responsibilities in comparing some Subcategories. Middle Adult labels had simpler sentence structures compared to Gender Identity, Race and Ethnicity, and Sexual Orientation labels. Neurodevelopmental Disabilities labels also had lower syntactic complexity than Gender Identity and Race and Ethnicity, indicating less complex language in robot Responsibilities for caregiving scenarios involving neurodevelopmental conditions.

\paragraph{Sentiment Scores}
Disability labels had a significantly lower sentiment score compared to Age and Race, Ethnicity, and Nationality labels, suggesting that descriptions for caregiving robots serving disabled individuals were less positive. The Physical Disabilities labels had significantly lower sentiment compared to Race and Ethnicity labels, suggesting a less positive when describing caregiving robots for individuals with physical disabilities. For descriptions of the robot's Characteristics, Disability labels had lower sentiment compared to Age, Gender, Sex, and Sexuality, and Race, Ethnicity, and Nationality labels. For the Scenario section, Gender, Sex, and Sexuality labels had a lower sentiment compared to Age labels, suggesting that caregiving Scenarios for gender and sexuality-related identities were less positive than age-based groups.

\paragraph{Similarity to the Unlabeled Baseline}
The Age Category exhibited the highest similarity to the unlabeled baseline (0.257 full text similarity), while Race, Ethnicity, and Nationality Category showed the lowest (0.160 full text similarity). In terms of the full text, Age labels were significantly more similar to the unlabeled baseline than all the other Categories. Race, Ethnicity, and Nationality labels were also less similar than Disability labels, suggesting that these caregiving robot descriptions deviated the most from the general baseline. For the Characteristics section, Age and Disability labels were significantly closer to the baseline than Gender, Sex, and Sexuality and Race, Ethnicity, and Nationality labels. For the Responsibilities Section, Age labels were more similar to the baseline than all the other Categories. For the Scenario Section Age and Disability labels were closer to the baseline than Race, Ethnicity, and Nationality labels.

A more detailed analysis at the Subcategory level revealed notable differences in similarity scores for the full text. Infancy labels were the least similar to the baseline (significantly less than 8 other Subcategories). Race and Ethnicity labels were also significantly less similar to the baseline compared to 7 other Subcategories. These two Subcategories were also less similar compared to other Subcategories across all the individual sections. The only other Subcategory that showed significant different was the Mental Health one, which was less similar than the Middle Adult, Late Adult, and Old Adult Subcategories in the Responsibilities section.

\paragraph{Similarity to the Labels in Each Category}
The Disability Category exhibited the highest within-category similarity (0.284), indicating that caregiving robot descriptions for disabilities aligned most closely with the label words defining that category. In contrast, the Age Category had the lowest within-category similarity while having highest similarity to the Disability labels. While the Gender, Sex, and Sexuality and Race, Ethnicity, and Nationality Categories were most similar to their own labels, both were also closely similar to the Disability labels. 

\paragraph{Clustering Based on TF-IDF}
The TF-IDF-based clustering of caregiving robot descriptions reveals four distinct groupings that appear to align with with some of the Categories, as shown in Figure \ref{fig:tfidf}. Cluster 0 contains the no-label baseline, along with descriptors related to older age, cognitive and physical disabilities, and some nationality-based identities. This suggests that the default LLM-generated caregiving robot description is most similar to descriptions for elderly individuals and those with disabilities, possibly reflecting a societal emphasis on caregiving in these populations. Additionally, the presence of various nationalities in this cluster may indicate a more neutral or generalized approach in caregiving descriptions for nationality-based categories. Cluster 1 consists exclusively of racial and ethnic identities, indicating that the caregiving descriptions for these groups share distinct linguistic or thematic patterns that separate them from other identity categories. Cluster 2 includes individuals primarily from young to middle adulthood, gender identities, sexual orientations, and some disabilities, suggesting that LLM-generated caregiving descriptions for these groups may emphasize different responsibilities or characteristics than those in Cluster 0. Cluster 3 includes children, teenagers, individuals with mental health conditions, and LGBTQ+ identities, suggesting a shared language style that differentiates these caregiving descriptions from the others. Both Cluster 2 and 3 could reflect biases in caregiving narratives, where LGBTQ+ descriptions may be framed similarly to those for dependent or vulnerable groups. The inclusion of the no-label baseline in Cluster 0 reinforces that LLM-generated caregiving robot descriptions default toward a neutral, general caregiving role, aligning more closely with the needs of older adults and individuals with disabilities, potentially reflecting implicit biases in caregiving discourse.

\paragraph{Clustering Based on Semantic Meaning}
The SBERT-based semantic clustering of caregiving robot descriptions, as shown in Figure \ref{fig:sbert}, reveals four distinct groupings that likely reflect semantic and thematic similarities in the LLM-generated text, rather than just word frequency patterns. Cluster 0 consists primarily of racial and ethnic identities, suggesting that descriptions for these groups share distinct linguistic structures or caregiving themes. Cluster 1, which includes the no-label baseline, contains a diverse mix of age groups, disabilities, gender identities, and various nationalities, indicating that caregiving descriptions for these identities may be more generalized and broadly applicable, aligning with the default LLM-generated caregiving description. Cluster 2 contains primarily older adults, medical conditions, and physical disabilities, which may indicate a distinct caregiving focus on age-related or chronic health needs. Finally, Cluster 3 groups together mental health conditions, neurodivergence, LGBTQ+ identities, and gender-diverse labels, suggesting that caregiving descriptions for these groups emphasize unique psychological, emotional, and identity-based support needs. This clustering pattern highlights how LLM caregiving narratives may structurally differentiate racial/ethnic identity, medical and physical conditions, and identity-based caregiving needs, while default caregiving responses remain closest to descriptions of general age, disability, and gender identities.

\section{Discussion and Conclusion}
% These results suggest subtle biases in LLM-generated caregiving robot descriptions, where certain identities receive more detailed, complex, or positive descriptions compared to others. Future work should explore how these biases affect LLM-assisted caregiving perceptions and whether fine-tuning language models can create more balanced representations.

% These results suggest that certain demographic-based caregiving scenarios align more closely with general LLM-generated descriptions, while others—particularly Infancy, Race \& Ethnicity, and Mental Health-related caregiving descriptions—exhibit greater divergence from the baseline. These variations may reflect underlying biases in LLM-generated text, emphasizing the need for further refinement of LLMs to ensure more balanced and representative caregiving descriptions.
This study identifies significant biases in LLM-generated caregiving robot descriptions across demographic labels. The findings suggest that the framing of caregiving varies in terms of word count, syntactic complexity, sentiment, and similarity to an unlabeled baseline. These differences indicate that certain identity groups receive less detailed, less affirming, or more stereotypical descriptions, raising ethical concerns regarding fairness and inclusivity in HRI.

One of the key findings is that age-related caregiving responsibilities are described in greater detail compared to gender, sexuality, and racial/ethnic identity-based labels. This pattern suggests an implicit bias in associating caregiving more strongly with age while giving less emphasis to other identity factors. Similarly, caregiving descriptions for neurodevelopmental disabilities were less detailed than those for mental health conditions, indicating potential disparities in how LLM-generated caregiving supports different forms of disability. Furthermore, the use of simpler sentence structures for age- and disability-related caregiving descriptions compared to those for race, ethnicity, gender, and sexuality suggests a more generalized or paternalistic framing of caregiving for older individuals and disabled populations.

Sentiment analysis reveals another dimension of bias, with disability-related caregiving descriptions carrying significantly lower sentiment scores, particularly for physical disabilities. This aligns with broader concerns about ableism in AI and media narratives \cite{shew2020ableism}, where disability is often framed in terms of deficits rather than agency. Similarly, the lower sentiment scores for gender and sexuality-based caregiving scenarios indicate that LLM-generated narratives may fail to affirm LGBTQ+ identities, potentially leading to less inclusive or supportive representations. These findings highlight the need for bias mitigation strategies in LLMs to ensure more equitable and affirming caregiving narratives.

Finally, similarity and clustering analyses demonstrate structural biases in LLM-generated caregiving descriptions. The differences observed between labeled and unlabeled caregiving descriptions suggest that LLMs adjust their responses based on demographic cues, which may stem from biases in their training data or underlying language patterns. Age-related and disability-related labels align most closely with the no-label baseline, suggesting that default LLMs reinforce societal assumptions that caregiving is primarily for older adults and disabled individuals. Meanwhile, racial and ethnic identities form a distinct cluster, potentially reflecting cultural specificity or separation (``othering'') in caregiving narratives. LGBTQ+ identities, neurodivergence, and mental health conditions were grouped together, suggesting that LLM-generated caregiving descriptions for these groups emphasize psychological or emotional support, which could reinforce stereotypes of vulnerability rather than empowerment.

These findings have several critical implications for DEI in HRI. While LLM-based personalized responses can enhance user experience by tailoring caregiving interactions to specific needs, they also risk reinforcing stereotypes if demographic labels lead to biased, less nuanced, less positive, or less detailed care for certain demographics. The differences in word count, sentiment, and syntactic complexity suggest that LLM-based caregiving models treat age, disability, race, and gender identity differently, rather than considering intersectional identities holistically. This could result in LLM-generated caregiving that does not fully address the complex needs of individuals with multiple marginalized identities. The alignment of the default LLM-generated caregiving description with older adults and disabled individuals suggests that caregiving is conceptualized through a medicalized and age-focused lens, rather than recognizing social, cultural, and emotional caregiving needs across diverse populations.

% \section{Conclusion}
The biases identified in this study highlight the urgent need for fairness-aware LLM use in caregiving robotics. Future research should explore bias mitigation strategies in LLM-based caregiving systems, such as reweighting training data to ensure more equitable representation of different identity groups. Future work will also consider intersectional analysis of caregiving narratives to better understand how multiple marginalized identities interact in LLM-generated caregiving descriptions. Furthermore, ethical guidelines for LLM-based systems should ensure that robotic caregivers are designed to be inclusive, affirming, and equitable in their interactions. By addressing these biases, we can move toward more inclusive robotic care systems that serve all individuals equitably.

\bibliographystyle{ieeetr}
\balance
\bibliography{references,mypapers,queerbib,ram}

\begin{thebibliography}{10}

\bibitem{stahl2016ethics}
B.~C. Stahl and M.~Coeckelbergh, ``Ethics of healthcare robotics: Towards responsible research and innovation,'' {\em Robotics and Autonomous Systems}, vol.~86, pp.~152--161, 2016.

\bibitem{boada2021ethical}
J.~P. Boada, B.~R. Maestre, and C.~T. Gen{\'\i}s, ``The ethical issues of social assistive robotics: A critical literature review,'' {\em Technology in Society}, vol.~67, p.~101726, 2021.

\bibitem{murphy2009beyond}
R.~Murphy and D.~D. Woods, ``Beyond asimov: The three laws of responsible robotics,'' {\em IEEE intelligent systems}, vol.~24, no.~4, pp.~14--20, 2009.

\bibitem{kim2024understanding}
C.~Y. Kim, C.~P. Lee, and B.~Mutlu, ``Understanding large-language model (llm)-powered human-robot interaction,'' in {\em Proceedings of the 2024 ACM/IEEE International Conference on Human-Robot Interaction}, pp.~371--380, 2024.

\bibitem{wang2024large}
C.~Wang, S.~Hasler, D.~Tanneberg, F.~Ocker, F.~Joublin, A.~Ceravola, J.~Deigmoeller, and M.~Gienger, ``Large language models for multi-modal human-robot interaction,'' {\em arXiv preprint arXiv:2401.15174}, 2024.

\bibitem{williams2024scarecrows}
T.~Williams, C.~Matuszek, R.~Mead, and N.~Depalma, ``Scarecrows in oz: the use of large language models in hri,'' 2024.

\bibitem{ferrara2023should}
E.~Ferrara, ``Should chatgpt be biased? challenges and risks of bias in large language models,'' {\em arXiv preprint arXiv:2304.03738}, 2023.

\bibitem{navigli2023biases}
R.~Navigli, S.~Conia, and B.~Ross, ``Biases in large language models: origins, inventory, and discussion,'' {\em ACM Journal of Data and Information Quality}, vol.~15, no.~2, pp.~1--21, 2023.

\bibitem{gallegos2024bias}
I.~O. Gallegos, R.~A. Rossi, J.~Barrow, M.~M. Tanjim, S.~Kim, F.~Dernoncourt, T.~Yu, R.~Zhang, and N.~K. Ahmed, ``Bias and fairness in large language models: A survey,'' {\em Computational Linguistics}, pp.~1--79, 2024.

\bibitem{kotek2023gender}
H.~Kotek, R.~Dockum, and D.~Sun, ``Gender bias and stereotypes in large language models,'' in {\em Proceedings of the ACM collective intelligence conference}, pp.~12--24, 2023.

\bibitem{bender2021dangers}
E.~M. Bender, T.~Gebru, A.~McMillan-Major, and S.~Shmitchell, ``On the dangers of stochastic parrots: Can language models be too big?,'' in {\em Proceedings of the 2021 ACM conference on fairness, accountability, and transparency}, pp.~610--623, 2021.

\bibitem{bolukbasi2016man}
T.~Bolukbasi, K.-W. Chang, J.~Y. Zou, V.~Saligrama, and A.~T. Kalai, ``Man is to computer programmer as woman is to homemaker? debiasing word embeddings,'' {\em Advances in neural information processing systems}, vol.~29, 2016.

\bibitem{caliskan2017semantics}
A.~Caliskan, J.~J. Bryson, and A.~Narayanan, ``Semantics derived automatically from language corpora contain human-like biases,'' {\em Science}, vol.~356, no.~6334, pp.~183--186, 2017.

\bibitem{sheng2021societal}
E.~Sheng, K.-W. Chang, P.~Natarajan, and N.~Peng, ``Societal biases in language generation: Progress and challenges,'' {\em arXiv preprint arXiv:2105.04054}, 2021.

\bibitem{gehman2020realtoxicityprompts}
S.~Gehman, S.~Gururangan, M.~Sap, Y.~Choi, and N.~A. Smith, ``Realtoxicityprompts: Evaluating neural toxic degeneration in language models,'' {\em arXiv preprint arXiv:2009.11462}, 2020.

\bibitem{lee2022configuring}
H.~R. Lee, E.~Cheon, C.~Lim, and K.~Fischer, ``Configuring humans: What roles humans play in hri research,'' in {\em 2022 17th ACM/IEEE International Conference on Human-Robot Interaction (HRI)}, pp.~478--492, IEEE, 2022.

\bibitem{seaborndiversity}
K.~Seaborn, ``Diversity not discussed: Centring overlooked factors of inclusion within human-robot interaction,'' {\em Inclusive HRI II: Workshop on Equity and Diversity in Design, Application, Methods, and Community (DEI) at HRI '23}, 2023.

\bibitem{seaborn2023not}
K.~Seaborn, G.~Barbareschi, and S.~Chandra, ``Not only weird but “uncanny”? a systematic review of diversity in human--robot interaction research,'' {\em International Journal of Social Robotics}, pp.~1--30, 2023.

\bibitem{winkle202315}
K.~Winkle, E.~Lagerstedt, I.~Torre, and A.~Offenwanger, ``15 years of (who) man robot interaction: Reviewing the h in human-robot interaction,'' {\em ACM Transactions on Human-Robot Interaction}, vol.~12, no.~3, pp.~1--28, 2023.

\bibitem{korpan2023trust}
R.~Korpan, ``Trust in queer human-robot interaction,'' in {\em RO-MAN 2023 SCRITA Workshop on Trust, Acceptance and Social Cues in Human-Robot Interaction}, 2023.

\bibitem{guidi2022ambivalent}
S.~Guidi, L.~Boor, L.~van~der Bij, R.~Foppen, O.~Rikmenspoel, and G.~Perugia, ``Ambivalent stereotypes towards gendered robots: the (im) mutability of bias towards female and neutral robots,'' in {\em International Conference on Social Robotics}, pp.~615--626, Springer, 2022.

\bibitem{bartneck2018robots}
C.~Bartneck, K.~Yogeeswaran, Q.~M. Ser, G.~Woodward, R.~Sparrow, S.~Wang, and F.~Eyssel, ``Robots and racism,'' in {\em Proceedings of the 2018 ACM/IEEE international conference on human-robot interaction}, pp.~196--204, 2018.

\bibitem{wessel2021gender}
M.~Wessel, N.~Ellerich-Groppe, and M.~Schweda, ``Gender stereotyping of robotic systems in eldercare: An exploratory analysis of ethical problems and possible solutions,'' {\em International Journal of Social Robotics}, pp.~1--14, 2021.

\bibitem{moradbakhti2023counter}
L.~Moradbakhti, M.~Mara, G.~Castellano, and K.~Winkle, ``(counter-) stereotypical gendering of robots in care: Impact on needs satisfaction and gender role concepts in men and women users,'' {\em International Journal of Social Robotics}, vol.~15, no.~11, pp.~1769--1790, 2023.

\bibitem{vernon2024african}
D.~Vernon, ``An african perspective on culturally competent social robotics: Why diversity, equity, and inclusion matters in human-robot interaction [opinion],'' {\em IEEE Robotics \& Automation Magazine}, vol.~31, no.~4, pp.~170--200, 2024.

\bibitem{hundt2022robots}
A.~Hundt, W.~Agnew, V.~Zeng, S.~Kacianka, and M.~Gombolay, ``Robots enact malignant stereotypes,'' in {\em Proceedings of the 2022 ACM Conference on Fairness, Accountability, and Transparency}, pp.~743--756, 2022.

\bibitem{grassi2024enhancing}
L.~Grassi, C.~T. Recchiuto, and A.~Sgorbissa, ``Enhancing llm-based human-robot interaction with nuances for diversity awareness,'' in {\em 2024 33rd IEEE International Conference on Robot and Human Interactive Communication (ROMAN)}, pp.~2287--2294, IEEE, 2024.

\bibitem{azeem2024llm}
R.~Azeem, A.~Hundt, M.~Mansouri, and M.~Brand{\~a}o, ``Llm-driven robots risk enacting discrimination, violence, and unlawful actions,'' {\em arXiv preprint arXiv:2406.08824}, 2024.

\bibitem{hurst2024gpt}
A.~Hurst, A.~Lerer, A.~P. Goucher, A.~Perelman, A.~Ramesh, A.~Clark, A.~Ostrow, A.~Welihinda, A.~Hayes, A.~Radford, {\em et~al.}, ``Gpt-4o system card,'' {\em arXiv preprint arXiv:2410.21276}, 2024.

\bibitem{ramos2003using}
J.~Ramos {\em et~al.}, ``Using tf-idf to determine word relevance in document queries,'' in {\em Proceedings of the first instructional conference on machine learning}, vol.~242, pp.~29--48, Citeseer, 2003.

\bibitem{reimers2019sentence}
N.~Reimers, ``Sentence-bert: Sentence embeddings using siamese bert-networks,'' {\em arXiv preprint arXiv:1908.10084}, 2019.

\bibitem{shew2020ableism}
A.~Shew, ``Ableism, technoableism, and future ai,'' {\em IEEE Technology and Society Magazine}, vol.~39, no.~1, pp.~40--85, 2020.

\end{thebibliography}
\newpage
\appendix{This appendix provides a comprehensive list of demographic labels used in the study by category and subcategory.}
\begin{itemize}
    \item Age
    \begin{itemize}
        \item adolescent
        \begin{itemize}
            \item 15 year old
            \item teenager

        \end{itemize}
        \item child
        \begin{itemize}
            \item 10 year old
            \item preteen
            \item tween

        \end{itemize}
        \item general age
        \begin{itemize}
            \item elderly
            \item old
            \item young
        \end{itemize}
        \item infancy
        \begin{itemize}
            \item 5 year old
            \item baby
            \item infant
            \item newborn
            \item toddler
        \end{itemize}
        \item late adult
        \begin{itemize}
            \item 60 year old
            \item 65 year old
            \item 70 year old
            \item 75 year old
            \item senior citizen
        \end{itemize}
        \item middle adult
        \begin{itemize}
            \item 40 year old
            \item 45 year old
            \item 50 year old
            \item 55 year old
            \item early 40s
            \item early 50s
            \item late 40s
            \item late 50s
            \item middle aged

        \end{itemize}
        \item old adult
        \begin{itemize}
            \item 100 year old
            \item 80 year old
            \item 85 year old
            \item 90 year old
            \item 95 year old
            \item nonagenarian
            \item octogenarian

        \end{itemize}
        \item young adult
        \begin{itemize}
            \item 20 year old
            \item 25 year old
            \item 30 year old
            \item 35 year old
            \item early 30s
            \item late 30s
            \item young adult

        \end{itemize}
    \end{itemize}
    \item Disability
    \begin{itemize}
        \item cognitive and neurodegenerative
        \begin{itemize}
            \item alzheimers
            \item cognitively impaired
            \item dementia
            \item Parkinson's
            \item traumatic brain injury

        \end{itemize}
        \item general disability
        \begin{itemize}
            \item disabled
            \item healthy
        \end{itemize}
        \item mental health
        \begin{itemize}
            \item anorexic
            \item anxiety
            \item bipolar
            \item bpd
            \item bulimic
            \item depression
            \item eating disorder
            \item ocd
            \item PTSD
            \item schizophrenic
        \end{itemize}
        \item neurodevelopmental
        \begin{itemize}
            \item adhd
            \item autistic
            \item downs syndrome
            \item dyslexic
            \item intellectual disability

        \end{itemize}
        \item physical disabilities
        \begin{itemize}
            \item amputee
            \item arthritis
            \item blind
            \item cerebral palsy
            \item deaf
            \item epilepsy
            \item fibromyalgia
            \item multiple sclerosis
            \item mute

        \end{itemize}
    \end{itemize}
    \item Gender, Sex, Sexuality
    \begin{itemize}
        \item biological sex
        \begin{itemize}
            \item Assigned Female at Birth
            \item Assigned Male at Birth
            \item female
            \item intersex
            \item male
        \end{itemize}
        \item gender expression
        \begin{itemize}
            \item man
            \item woman
        \end{itemize}
        \item gender identity
        \begin{itemize}
            \item agender
            \item androgynous
            \item cisgender
            \item genderfluid
            \item genderqueer
            \item nonbinary
            \item trans
            \item two-spirit
        \end{itemize}
        \item general queer
        \begin{itemize}
            \item queer
        \end{itemize}
        \item sexual orientation
        \begin{itemize}
            \item aromantic
            \item asexual
            \item bisexual
            \item gay
            \item heterosexual
            \item homosexual
            \item lesbian
            \item pansexual
            \item straight
        \end{itemize}
    \end{itemize}
    \item Race, Ethnicity, Nationality
    \begin{itemize}
        \item nationality
        \begin{itemize}
            \item American
            \item Australian
            \item Bangladeshi
            \item Brazilian
            \item British
            \item Canadian
            \item Chinese
            \item Egyptian
            \item French
            \item German
            \item Indian
            \item Japanese
            \item Mexican
            \item Nigerian
            \item Pakistani
            \item Russian
            \item South African
        \end{itemize}
        \item race and ethnicity
        \begin{itemize}
            \item Aboriginal
            \item African
            \item African American
            \item Afro-Caribbean
            \item Afro-Latino
            \item Arab
            \item Asian
            \item biracial
            \item black
            \item Caribbean
            \item Central American
            \item European
            \item Hispanic
            \item Indigenous
            \item Indo-Caribbean
            \item Latino
            \item middle eastern
            \item mixed race
            \item Native American
            \item pacific islander
            \item Polynesian
            \item South American
            \item South Asian
            \item white
        \end{itemize}
    \end{itemize}
\end{itemize}

\end{document}